\documentstyle[12pt,epsf]{article}
\newlength{\dinwidth}
\newlength{\dinmargin}
\begin{document}
\input{psfig}
\newcommand{\ra}{\rightarrow}
\newcommand{\as}{\mbox{$\alpha_{\displaystyle  s}$}}
\vspace{1 cm}
\title {
{\bf Observation of hard scattering in photoproduction events with a large
rapidity gap at HERA}
     \\
\author{ ZEUS COLLABORATION\\
                       } }
\date{ }
% {\it FINAL VERSION}}
\maketitle
%---------------
\def\pom{{\cal P}}
\def\reg{{\cal R}}
\def\sm0{{ o}}
\def\pt{{\rm \P_t}}
\def\pbarp{{\rm \overline{p}p}}
\def\mev{{\ \rm MeV}}
\def\gev{{\rm\  GeV}}
\def\fg#1{\noindent { figure #1}}
\def\f2g{$F_2^{\gamma}$}
\vspace{5 cm}
\begin{abstract}
Events with a large rapidity gap and total transverse energy greater than 5 GeV
have been observed in quasi-real photoproduction at HERA
with the ZEUS detector. The distribution of these events as a function
of the $\gamma p$ centre of mass energy is consistent with diffractive
scattering. For total
transverse energies above 12 GeV, the hadronic final states show
predominantly a two-jet structure with each jet
having a transverse energy greater than 4 GeV.
For the two-jet events, little energy flow is found outside the jets. This
observation is consistent with the hard
scattering of a quasi-real photon with a colourless object in the
proton.
\\
\end{abstract}
\vspace{-20cm}
\begin{flushleft}
\tt DESY 94-210 \\
November 1994 \\
\end{flushleft}
\setcounter{page}{0}
\thispagestyle{empty}   % to suppress the page number on the first page
\vspace{3 cm}
\newpage

\def\3{\ss}
\footnotesize
\renewcommand{\thepage}{\Roman{page}}
\begin{center}
\begin{large}
The ZEUS Collaboration
\end{large}
\end{center}
\noindent
M.~Derrick, D.~Krakauer, S.~Magill, B.~Musgrave, J.~Repond,
J.~Schlereth, R.~Stanek, R.L.~Talaga, J.~Thron \\
{\it Argonne National Laboratory, Argonne, IL, USA}~$^{p}$\\[6pt]
F.~Arzarello
% ...
\\[6pt]
\clearpage
\noindent
\hspace*{1mm}
% ...

\begin{tabular}{lp{15cm}}
$^{a}$ &supported by the Natural Sciences and Engineering Research
         Council of Canada \\
$^{b}$ &supported by the FCAR of Quebec, Canada\\
$^{c}$ &supported by the German Federal Ministry for Research and
         Technology (BMFT)\\
$^{d}$ &supported by the MINERVA Gesellschaft f\"ur Forschung GmbH,
         and by the Israel Academy of Science \\
$^{e}$ &supported by the German Israeli Foundation, and
         by the Israel Academy of Science \\
$^{f}$ &supported by the Italian National Institute for Nuclear Physics
         (INFN) \\
$^{g}$ &supported by the Japanese Ministry of Education, Science and
         Culture (the Monbusho)
         and its grants for Scientific Research\\
$^{h}$ &supported by the Korean Ministry of Education and Korea Science
         and Engineering Foundation \\
$^{i}$ &supported by the Netherlands Foundation for Research on Matter
         (FOM)\\
$^{j}$ &supported by the Polish State Committee for Scientific
         Research (grant No. 204209101) \\
$^{k}$ &supported by the Polish State Committee for Scientific Research
         (grant No. PB 861/2/91 and No. 2 2372 9102,
         grant No. PB 2 2376 9102 and No. PB 2 0092 9101) \\
$^{l}$ &partially supported by the German Federal Ministry for
         Research and Technology (BMFT) \\
$^{m}$ &supported by the German Federal Ministry for Research and
         Technology (BMFT), the Volkswagen Foundation, and the Deutsche
         Forschungsgemeinschaft \\
$^{n}$ &supported by the Spanish Ministry of Education and Science
         through funds provided by CICYT \\
$^{o}$ &supported by the Particle Physics and Astronomy Research
        Council \\
$^{p}$ &supported by the US Department of Energy \\
$^{q}$ &supported by the US National Science Foundation
\end{tabular}

\newpage
\pagenumbering{arabic}
\setcounter{page}{1}
\normalsize

\section{\bf Introduction}

\par
In a recent publication \cite{sigtot}, it has been shown that in
photoproduction
at HERA energies the contribution of
diffractive processes account for about 36\% of the total photoproduction cross
section. Diffractive processes are generally believed to proceed via the
t-channel exchange of a colour-singlet object,
with vacuum quantum numbers and which carries energy-momentum,
called the pomeron. The true nature of
the pomeron is still far from clear.  Ingelman and Schlein \cite{gips}
assumed that the pomeron emitted from the proton behaves like a hadron
and suggested that it could have a partonic
substructure which could be probed by a hard
scattering process. The UA8 experiment at CERN later observed events containing
two high-$p_T$ jets in $\overline{p}p$ interactions tagged with
leading protons (or antiprotons) \cite{ua8}.
These observations could be explained in terms of a partonic structure
in the pomeron.

\par
In previous publications \cite{zlrg1, zlrg2,zlrg3,h1lrg}
evidence has been presented for events with a large rapidity gap
in deep inelastic scattering (DIS). The energy dependence of the event rate
as well as the approximate scaling  present in the data
pointed to a diffractive process of a leading twist
nature. The hadronic final state of these events
exhibited a small, but significant, rate for two-jet
production in the $\gamma^* p$ frame. The characteristics
of these events are consistent with an interaction between a
virtual photon and  partons in a colourless object
from the proton.
In this context the term `pomeron exchange'
is used as a generic name to describe the
process which is responsible for creating events with a large rapidity gap.

In this paper we report the observation of events with a large rapidity gap
in a sample of events
with high transverse energy produced in the photoproduction regime
($Q^2$ $\approx0$, where $-Q^2$ is the four-momentum transfer squared carried
by the virtual photon).
The analysis of the associated hadronic final states in these events includes
the study of jet
structure to search for a hard scattering process in diffractive
photoproduction.

\section{Experimental setup}
\subsection{HERA machine conditions}

\par
The data presented here were obtained with the ZEUS detector during the 1993
running period at the electron-proton collider HERA when 84
bunches of electrons with energy $E_e=26.7$ GeV collided
with 84 bunches of protons of energy $E_p$= 820 GeV.
In addition, 10 electron and 6 proton non-colliding bunches were used
for studies of beam induced background.
The electron and proton beam
currents were typically 10 mA.

\subsection{The ZEUS detector}

ZEUS is a multipurpose magnetic detector whose configuration has been
described elsewhere \cite{sigtot, zlrg2}. For the present analysis we
used only some of the components within ZEUS.
Charged particles are tracked by the vertex detector
(VXD) \cite{vxd} and the central tracking detector (CTD) \cite{ctd}
which operate inside a thin superconducting solenoid providing an axial
magnetic field of 1.43 T . The solenoid is surrounded by a high
resolution uranium-scintillator
calorimeter divided into three parts,
forward (FCAL) covering the pseudorapidity \footnote{The ZEUS coordinate
system is defined
 as right-handed with the $Z$ axis pointing in the (forward)
proton direction and the $X$
axis pointing horizontally towards the centre of the HERA rings.
The pseudorapidity
$\eta$ is defined as $-\ln(\tan \frac{\theta}{2})$, where the polar
angle $\theta$ is taken with respect to the proton beam direction from
the nominal interaction point (IP).} region $4.3 \geq \eta \geq 1.1$, barrel
(BCAL) covering
the central region $1.1 \geq \eta \geq -0.75$ and rear (RCAL) covering
the backward region $-0.75 \geq \eta \geq -3.8$. The resulting
solid angle coverage is $99.7\%$ of $4\pi$.
The calorimeter parts are subdivided into towers which in turn are subdivided
longitudinally into electromagnetic (EMC) and hadronic (HAC) sections. The
sections are subdivided into cells,  each of which is viewed by two
photomultiplier tubes.
The calorimeter is described in detail elsewhere \cite{test1}.
For measuring the luminosity as well as for tagging the scattered electron
in  small $Q^2$
processes, we use two lead-scintillator calorimeters \cite{lumi}. For
these `tagged' photoproduction events, the resulting $Q^2$ values
are less than 0.02 GeV$^2$.

\subsection{Trigger conditions}

The data were
collected with a three-level trigger.
The First Level Trigger (FLT), based on a deadtime-free pipeline,
selects inclusive photoproduction events with a calorimeter energy
trigger. For FLT purposes, the calorimeter is subdivided
into 896 trigger towers, each tower consisting of an EMC
and a HAC segment. Events for this analysis were accepted by the FLT
if the energy sum of EMC towers in the BCAL exceeded 3.4 GeV, or in
the RCAL (excluding the cells adjacent to the beam pipe) exceeded
2.0 GeV, or the EMC towers  exceeded 3.75 GeV in the RCAL towers
adjacent to the beam-pipe.

\par
The Second Level Trigger (SLT) used information from a subset of detector
components to differentiate physics events from backgrounds.
The SLT rejected proton beam-gas background by
timing measurements in the calorimeter cells; this
algorithm reduced the RCAL and BCAL FLT rates by
approximately $90\%$ and $50\%$, respectively.

The Third Level Trigger (TLT) had available the full event information
 on which to apply physics-based filters. The TLT
 applied stricter cuts on the event times and also rejected beam-halo
muons and cosmic muons. The logic of the filter used in this analysis is
described in the data selection section.

\section{Kinematics of photoproduction events}

\par
In electron-proton scattering,
photoproduction can be studied in the limit of small four-mo-\newline mentum
transfer, $q$,  carried by the virtual photon, $\gamma^*$. The kinematic
variables used to describe the reaction
   $ e~(k)~+~ p~(P) \rightarrow e~(k')~+~X$, are the following:
the square of the total $ep$ centre of mass energy:$~s = (k~+~P)^2 \approx
4 E_p  E_e
= 87576 $ GeV$^2$;
the four-momentum transfer squared carried by the virtual photon:
$ Q^2=-q^2=-(k~-~k')^2 $;
the Bjorken variable describing the energy transfer to the hadronic system:
$y =\frac{q\cdot P}{k\cdot q}$;
and the centre of mass energy squared of the $\gamma^*p$ system:
$W^2=(q+P)^2  = ys$.
\par
For events with a detected electron,
the variable $y$ can be obtained from
\begin{equation}
y_e = 1 - \frac{E^{\prime}_e}{E_e}~\frac{1-cos \theta^{\prime}_e}{2}
\end{equation}
where $E^{\prime}_e$ denotes the scattered electron
energy and $\theta^{\prime}_e$ the electron scattering angle.
Alternatively, $y$ can be determined  approximately
from the hadronic system using the Jacquet-Blondel expression
\begin{equation}
 y_{JB}=\frac{\sum_i (E_i - p_{zi})}{2\cdot E_e}
\end{equation}
with the sum running over all calorimeter cells $i$ associated with the
hadronic system. We denote by $E_i$ the energy in cell
$i$, $p_{zi}=E_i\cdot cos\theta_i$ and $\theta_i$ is the angle of the
centre of the cell with respect to the event vertex.
The invariant mass, $M_X$, of the hadronic system
detected in the ZEUS central detector
is determined from the calorimeter cell information: $M_X^2=E_H^2 - p_H^2$,
where $E_H$ and $p_H$ are the energy and momentum
of the hadronic system. Cells in the electromagnetic
(hadronic) calorimeter sections with
energies below 60 MeV (110 MeV) were excluded in the present analysis.

\section{Data selection}

The cuts to select hard photoproduction events
are similar to those described in our
previous publications \cite{juan,zhpp}. In particular, we

\begin{itemize}
\item reject events with an electron found in the calorimeter with an
energy greater than 5 GeV and $y_e \leq 0.7$, to remove DIS background;
\item require $0.05\leq y_{JB} \leq 0.8$, to remove beam-gas and those DIS
events where the scattered electron was not identified and therefore was
mistakenly included in the hadronic system;
\item reject events with an energy in the rear section of the
calorimeter $E_{RCAL} > 30$ GeV, to remove background from non-ep collisions;
\item reject events with a missing transverse momentum, $\not p_T > 10$ GeV/c;
\item require a vertex with $-35 \leq Z \leq 20$ cm and
a radial distance from the beam line $R\leq 4$ cm;
\item require $\frac{\sum_i p_{zi}}{\sum_i E_i}\leq~0.9$,
at least two oppositely charged tracks with $p_T~\geq~0.5$ GeV, and
one of the following: transverse energy in a cone outside $10^{o}$
of the forward direction in excess of 5 GeV; or $y_{JB} \geq 0.28$;
or $\frac{\sum_i p_{zi}}{\sum_i E_i}~ \leq~0.8$; or an electron with
an energy larger
than 5 GeV detected in the luminosity monitor; these criteria
were essential to
enable us to reach lower transverse energies than in previous
publications \cite{juan};
\item require a total transverse energy $E_T \geq 5$ GeV.
\end{itemize}
\par
The last condition in particular defines `hard photoproduction' in the
context of this analysis.
\par
{}From an integrated luminosity of 0.55 pb$^{-1}$,
a sample of 417081 events passed these cuts.
This sample has less than a $0.1\%$ contamination from beam-gas
interactions, as determined from the number of events originating from
non-colliding electron
and proton bunches. The cosmic ray background, estimated from the rate of
events outside of $ep$  bunch crossings, is negligible.
 The requirement that no electron is found in the ZEUS calorimeter ensures
 that $Q^2 \leq 4$ GeV$^2$.
Monte Carlo studies, using the ALLM \cite{ALLM} prescription
which provides a smooth interpolation from deep inelastic
scattering to $Q^2$ = 0,
show that for the accepted photoproduction events
the median $Q^2$ is $10^{-3}$ GeV$^2$. The same Monte Carlo program
predicts that $27\%$ of the events should have a scattered electron
measured in the electron calorimeter of the luminosity detector, in
agreement with the observed fraction of $26.7\%$.

\section{The Monte-Carlo simulation }

The hadronic final states observed in hard photoproduction
can be understood
as the result of two different leading order (LO) mechanisms:
\begin{itemize}
\item direct processes where the photon interacts with a gluon ($g$) in
the proton giving rise to a quark-antiquark pair (Boson Gluon Fusion)
 or with a quark ($q$), generating a $qg$ final state (QCD
Compton), and
\item resolved processes \cite{witten} where a parton in the photon
interacts with a parton in the proton.
\end{itemize}

\par
Calculations of the relative importance of these
two competing mechanisms at HERA energies have shown that
the hard
photoproduction cross section is dominated by resolved processes
\cite{dre}. These expectations have been recently confirmed
experimentally \cite{juan,zhpp,h11}.

The Monte Carlo generator PYTHIA 5.6~\cite{p} was used to model
standard hard photoproduction processes. In this generator,
the direct and resolved photon processes are each simulated using
leading order matrix elements, with the inclusion of initial and
final state parton showers. The lower cut-off on the transverse
momentum of the generated final-state partons $p_{tmin}$ was chosen
to be 2.5 GeV/c. The
photon parton distributions were parametrized
according to GRV-LO \cite{GRV} while for the proton
MRSD$_{-}^\prime$ \cite{martin} was used. The
Weizs\"acker-Williams approximation was used to describe the photon flux at the
lepton vertex.
This Monte Carlo simulation does not contain any explicit contribution from
diffractive $\gamma$p interactions.

\par
Diffractive processes were simulated using POMPYT \cite{pompyt}.
This is a Monte Carlo model in which, within the framework
provided by PYTHIA, the proton emits a pomeron whose
partonic constituents subsequently take part in a hard scattering process
with the photon or its constituents. This model incorporates approximately
energy independent cross sections as experimentally determined in
hadron-hadron collisions.

The photon contribution contains both direct and resolved processes and
the parton density distribution
is parameterised according to DG \cite{DG}.
The parton momentum densities of the pomeron are
parameterised according to the hard distribution \cite{gips,pompyt}
\begin{equation}
\beta f(\beta)=constant \cdot~\beta(1-\beta)
\end{equation}
where $\beta=x_{parton/pom}$  denotes the fraction of the
pomeron momentum involved in the scattering.\footnote{The soft parton density
has not been considered since
it does not describe our DIS data \cite{zlrg2}.}
Two possibilities have been considered: one in which the partons in the
pomeron are quarks (quarkonic pomeron) and a second one in which
the partons are gluons (gluonic pomeron).
 The two predictions from the POMPYT model will be shown separately.

\par
In our previous publication on the study of jet production
in DIS events with a large rapidity gap, the Nikolaev-Zakharov (N-Z) model
\cite{nz} gave similar results to POMPYT. For the photoproduction processes
studied here, the N-Z model again gives similar results to POMPYT.

For all of the comparisons shown below, the number of Monte Carlo events
has been normalised to the data in each figure
and so only the shapes of the distributions
may be compared. The Monte Carlo events were passed through  reconstruction
and selection procedures identical to those for the data.

%All Monte Carlo events were passed through the standard ZEUS detector
%and trigger simulation as well as event selection and reconstruction programs.

\section{\bf Results}

\subsection{Events with large rapidity gaps}

\par
Following \cite{zlrg1, zlrg2} we define $\eta_{max}$ as the maximum
pseudorapidity
of all calorimeter condensates in an event, where
a condensate is defined as an isolated set of adjacent cells with summed
energy above 400 MeV. The pseudorapidity of the condensate is then
calculated from the angle of the energy weighted centre of the condensate
with respect to the measured IP.
The distribution of $\eta_{max}$ is shown in
Fig 1a.
The data presented in this and all following figures are not corrected
for effects from
detector acceptance and smearing. The dip in the $\eta_{max}$ distribution
at $\eta_{max} \approx $ 1.1 is a detector effect.
Values of $\eta_{max}\geq 4.3$, which are outside the acceptance of the
calorimeter, occur when energy is deposited in many contiguous
cells around the beam pipe in the forward (proton) direction. This
region is sensitive to the fragmentation of the proton remnant which at
 HERA energies is not yet fully understood. The bulk of the
events cluster around $\eta_{max} \sim 4$ in fair agreement with
the expectations of PYTHIA. In addition to this
region of large $\eta_{max}$ a second class of events with
$\eta_{max} \leq 1.5$ is seen in the data. There are
6678 events with $\eta_{max} \le $1.5, corresponding to 1.6 \% of the
total sample. The beam-gas contamination in this subsample is less than $1\%$
and that due to cosmic rays is 1.3\%.
%is kept at a similar level as that discussed for the complete
%sample in the preceeding section.

The shape of the distribution for $\eta_{max} \leq 1.5$
is not accounted for by standard Monte Carlo simulations
for hard photoproduction processes as in PYTHIA.
It is, however, in good agreement with the
predictions of the POMPYT model, as
illustrated in Fig. 1a.
Note that the normalisations of the
POMPYT and PYTHIA samples in this figure
have been fixed to the number of data events
below and above $\eta_{max}=1.5$, respectively.
According to POMPYT, for events
with $E_T > 5$ GeV the acceptance after the $\eta_{max} \le$ 1.5 cut is
about 9\% (10\%) for a quarkonic (gluonic) pomeron.

\par
In Regge phenomenology for soft processes the amplitudes for two-body
scattering by pomeron exchange are characterized by a constant or slowly rising
dependence on $W$, the centre of mass energy, while reggeon exchange
leads to a power law decrease. Hence, the $W$ dependence of the rate of large
rapidity gap events is a sensitive measure of the type of exchange
contributing to the scattering process. It is important to study whether
the same phenomenology exists for hard scattering, the subject of this
paper. The
distributions in $W=(y_{JB} \cdot s)^{\frac{1}{2}}$ for all events and for
those with $\eta_{max} \le $1.5
are shown in Fig. 1b.  Also shown are the POMPYT predictions for events
with $\eta_{max}\leq 1.5$. The good agreement of POMPYT with the data is
consistent with the assumption that the
dominant mechanism for large rapidity gap events is pomeron exchange.

\par
The mass of the hadronic system was measured with the calorimeter
as described at the end of section 3.
The distribution of $M_X$ for events with a large rapidity gap
is shown in Fig. 1c. The resolution in the determination of $M_X$ is between
10-20\%. According to Monte Carlo calculations the reconstructed $M_X$
values are
 typically underestimated by about 20\% with respect to the generated masses.
The $M_X$ distribution is steeply falling for $M_X$ values
above $12$ GeV. This is similar to
the behaviour observed in DIS for events selected with similar cuts
\cite{zlrg2}.
Neither the gluonic nor quarkonic POMPYT model gives a
satisfactory description of the data. The correlation between $M_X$ and
$\eta_{max}$ is displayed in Fig. 1d.  Note that, according to POMPYT,
there are
diffractive events at high $M_X$ values which are suppressed
by the  $\eta_{max}$ cut.

\par
If large rapidity gap events are interpreted as being due to pomeron
exchange, then the fraction of the proton momentum carried by the
pomeron $x_{pom/p}$ can be determined from the mass of the hadronic system
 via the relation
\begin{equation}
 x_{pom/p} = \frac{M_X^2}{W^2}.
\end{equation}
 For events with
$\eta_{max} \leq 1.5$, $x_{pom/p}$ clusters (not shown)
around values of $\sim 3\cdot 10^{-3}$ due to the applied cuts. The
$x_{pom/p}$ values are limited on the lower side to 7$\cdot 10^{-4} $
by the $E_T\geq5$ GeV requirement
while the $\eta_{max} \leq 1.5$ cut suppresses values of
$x_{pom/p}$ above
$10^{-2}$.
Similar $x_{pom/p}$ values have been observed in DIS \cite{zlrg2}.

\subsection{Jet structure }

Evidence for multijet structure in hard photoproduction at HERA has been
presented in \cite{juan, zhpp,h11}.
In the events with
$E_T \geq 5~$GeV a search was performed for jet structure using
a cone-based jet algorithm in
pseudorapidity ($\eta$), azimuth ($\phi$) space \cite{ua1}, subject to
the Snowmass convention \cite{huth}. The cone radius
$R=(\Delta \phi^2 + \Delta \eta ^2)^{\frac{1}{2}}$
in the algorithm was set to 1 unit. In order to ensure that for
standard (non-diffractive) hard photoproduction the results are not
biased by fragments from the proton remnant, whose fragmentation
properties at these energies are not well known, calorimeter cells
with polar angles smaller than $9^{\circ}$ ($\eta\geq2.5$) in the
laboratory were excluded.

In the first step of the jet algorithm, each calorimeter cell with a
transverse energy in excess of 300 MeV is considered as a seed for the search.
These seeds are combined if their distance in $\eta - \phi$ space, $R$, is
smaller than 1 unit. Then a cone of radius $R$ = 1 is drawn around each
seed and the calorimeter cells within that cone are combined to form a
cluster. The axis of the cluster is defined according to the Snowmass
convention: $\eta^{cluster} ~(\phi^{cluster})$ is the transverse energy
weighted mean pseudorapidity (azimuth) of all the calorimeter cells belonging
to that cluster. A new cone of radius 1 unit is then drawn around the axis of
the cluster. All cells inside the cone are used to recalculate a new cluster
axis. The procedure is iterated until the content of the cluster does not
change.

The energy sharing of overlapping clusters is then considered. Two clusters are
merged if the common transverse energy exceeds 75\% of the total transverse
energy of the cluster with the lowest transverse enrgy; otherwise two different
clusters are formed and the common cells are assigned to the nearest cluster.
Finally, a cluster is called a jet if $E_{T}^{cluster} \geq ~ 4$ GeV.
The transverse energy weighted mean pseudorapidity
$(\eta_{jet})$ and azimuth $(\phi_{jet})$ were evaluated and jets
were accepted for the present analysis if $\eta_{jet} \leq 2$, corresponding
to polar angles larger than $15^{\circ}$.

We find that 91.4\%, 6.5\%, 2.0\% and 0.1\%
of the events with total transverse energy E$_T >$ 5 GeV belong to the
zero-, one-, two- and three- or more-jet categories in the
sample of events with $\eta_{max}\leq1.5$. The fraction of
zero-jet events depends strongly on the jet definition, in particular on
the requirement that the jet transverse energy be larger than 4 GeV.

To determine the background present in the two-jet sample with a large
rapidity gap,
 we studied the empty as well as the non-colliding electron and proton bunches
and set an upper limit of
1\% for the beam-gas and cosmic ray backgrounds. For further confirmation
that this sample of
two-jet events is from photoproduction processes, we measured the fraction
of two-jet events
where the scattered electron is tagged in the electron calorimeter of the
luminosity monitor with an energy between $5$ and
$22$ GeV and for which $Q^2 < $ 0.02 GeV$^2$. This was found to be 24\%
in agreement with
both Monte Carlo expectations and the value for the complete high $E_T$
photoproduction sample.

\par
Figure 2a displays the $\eta_{max}$ distribution for all events with two or
more jets. A sample of 132 two-jet
events with $\eta_{max} \le 1.5$ is observed, corresponding to 0.63\% of
the hard photoproduction two-jet sample. This
number of events is not accounted for by the
standard hard photoproduction processes as modelled by PYTHIA which predicts
fewer than 18 $\pm$ 7 two-jet events with $\eta_{max} \leq$ 1.5.
 Therefore, the two-jet events observed with $\eta_{max} \le$ 1.5
are not just the tail
of `standard' hard photoproduction with two jets \cite{juan}.
Both POMPYT samples give a good representation of the shape of the
$\eta_{max}$ distribution for these  events.

For events with $\eta_{max}\leq 1.5$,
the distribution of the total transverse energy per event, $E_T$,
is shown in Fig. 2b for all events (open histogram), for events
with at least one jet in the final state (cross hatched) and for those
events with two or more jets (shaded). For $E_T \geq 8$ GeV, which is the
minimum $E_T$ for which two-jet production with $E_T \geq 4$ GeV
 is possible, $9.9\%$ of the
events are of the two-jet type. For $E_T \geq 12$ GeV the majority
of the events are of the two-jet type; an example
of which is shown in Fig. 2c.
For two-jet events the
distribution of the transverse jet energies reaches up to values
of 10 GeV, as shown in Fig. 3a.
In Fig. 3b the difference in azimuth $(\Delta \phi^{jet})$
of the two jets in the large rapidity
gap sample is displayed. The two jets are preferentially back-to-back in
the transverse plane. The predictions from both POMPYT models are in    fair
agreement with these data.

\subsection{Transverse energy flow around the jet axis}

The profiles of the jets observed in
high $E_T$ photoproduction events are compared for events with
and without a large rapidity gap. Figures 3c, d show the transverse energy
flows around the jet axis.
To reduce the bias from the $\eta_{max} \le 1.5$ cut,
the jets for both samples were restricted to the region $\eta_{jet} \leq $0
and calorimeter cell energy deposits with $\eta_{cell} >$ 1.5 were excluded.
Fig 3c shows that the $E_T$-weighted
azimuthal distributions ($\Delta \phi$ = $\phi_{cell}
-\phi_{jet~axis}$) in the jet core are essentially
the same for the large rapidity gap events ($\eta_{max} \le 1.5$)
and for the hard photoproduction sample with $\eta_{max} > 1.5$. However, the
transverse energy outside of the jet core is about a factor of two larger
for the $\eta_{max} > 1.5$ sample than for the large rapidity gap events.

Figure 3d shows the $E_T$-weighted distribution for
$\Delta \eta$, the difference in
rapidity of a given calorimeter cell and the jet axis. For this figure, only
energy deposits in the hemisphere containing the jet are included.
The transverse energy flow in the core of the jet is the same for events
with and without a large rapidity gap, indicating that the jets themselves
are similar. However, jets produced in events without a large
rapidity gap show at large (forward) $\Delta \eta$
values a significant $E_T$ flow.
In contrast, the jet profile for the large rapidity gap sample is more
symmetric and shows a factor of two to three less $E_T$ flow in the
forward direction. These observations are similar to those
presented in our previous publications
on DIS events with and without a
large rapidity gap \cite{zlrg2,zlrg3} where it was
demonstrated there is a suppression of colour flow
between the outgoing nucleon system and the struck parton in events with a
large rapidity gap.

\subsection{Momentum fractions}

 We conclude that we have observed large rapidity gap events containing
two jets,
consistent with a photon pomeron hard scattering
process leading to a two-jet final state.
In $2\rightarrow 2$ parton scattering the momenta of the incoming
partons can be calculated from the two partons in the final state. Let
$x_{\gamma}$ and $x_p$ be the fractions of the photon and proton
momenta carried by the initial state partons. If the final state partons are
approximated by the two jets observed, energy-momentum
conservation leads to the relationship:
\begin{equation}
 x_{\gamma}= \frac{\sum (E-p_z)_{Jets}}{\sum (E-p_z)_{hadrons}} ,
\end{equation}
\begin{equation}
 x_p= \frac{\sum (E+p_z)_{Jets}}{2E_p}.
\end{equation}
In these expressions `Jets' refers to a sum over all calorimeter cells
that comprise the jets according to the cone algorithm employed.
Direct photon processes are characterised by $x_{\gamma}$ values
close to 1.

Equations (5) and (6) were used to calculate $x_\gamma$ and $x_p$. The
resolutions are
comparable to those reported in an earlier publication on the hard
photoproduction of two jets\cite{juan} and are in the range (10-20)\%.
 The $x_\gamma$ distribution
shown in Fig. 4a peaks near $x_\gamma=1$. The sum of the direct and
resolved photon contributions as calculated by POMPYT for a
gluonic pomeron gives a reasonable description of the
data, as does a pure direct photon interaction with a quarkonic pomeron.
In contrast, the shape of a purely resolved photon contribution
is inadequate to describe the data, as seen by  the
dashed-dotted histogram in Fig.
4a which was calculated for a quarkonic pomeron with Eq. (3).

 The distribution of $x_p$,
the fraction of the proton's
momentum participating in the hard scattering, is displayed in Fig 4b and shows
that most of the two-jet events have $x_p$ values between 3$\cdot 10^{-3}$ and
$10^{-2}$. The low and high limits are due to the $E_{T}$ and $\eta_{max}$
cuts, respectively, as discussed in section 6.1.
In this sample, the two jets populate the pseudorapidity
region $-2 \leq \eta_{jet} \leq 1$ as shown
in Fig. 4c, because
the $\eta_{max} \leq 1.5$ cut suppresses jets with $\eta_{jet} > 1$.

In standard hard photoproduction at HERA, the centre-of-mass of the hard
process is boosted in the proton direction and, therefore, in most
two-jet events both jets go forward~\cite{juan}. This is particularly true
for the resolved photon process which is, in general, the dominant jet
production process at the transverse energies considered here.
At centre-of-mass
energies of $W \sim 200$ GeV, the direct photon process is expected to
dominate only at transverse energies above 50 GeV \cite{dre}. In the
selected sample of large rapidity gap events,
 $M_X$ is an order of magnitude smaller than $W$, thus
limiting the available phase space. The
dominance of the direct component hence
occurs at transverse energies an order of
magnitude smaller than in standard hard photoproduction.
Therefore the fact that two-jet events in the large
rapidity gap sample are found to be mainly due to the direct
coupling of the photon is well understood with the conventional
parton density parametrisations of the photon.
These features are reproduced by the POMPYT model.

\section{Conclusions}

\par
We have observed photoproduction events with a large rapidity gap
and large transverse energy at HERA.
Their distribution
as a function of the $\gamma p$ centre of mass energy is consistent
with a diffractive process. Hard scattering, with jets having transverse
energies greater than 4 GeV, has been observed in these large rapidity gap
events. For total transverse energies above 12 GeV the
hadronic final state is dominated by two-jet production with the two jets
being preferentially back-to-back in azimuth. For the two-jet events
selected by the cuts used in this analysis, the fraction of the photon
momentum participating in the hard scattering is close to one, suggesting that
their production is dominated by direct photon processes.
The two-jet events in the
large rapidity gap sample show little energy outside the jet core. A
natural interpretation of these events is the hard interaction of the
 photon with a colourless object inside the proton: the pomeron.
These conclusions complement those drawn from the production of
large rapidity gap events in deep inelastic scattering.

\vspace{2cm}

\noindent {\Large\bf Acknowledgements}

We thank the DESY Directorate for their strong support and encouragement.
The remarkable achievements of the HERA machine group were essential for the
successful completion of this work and we are grateful for their efforts
which enabled us to achieve a factor of twenty increase in the integrated
luminosity for the 1993 running period over that achieved in 1992.

%--------- REFERENCES -------------

% fig 1

\newpage
\begin{figure}[h]
\includegraphics{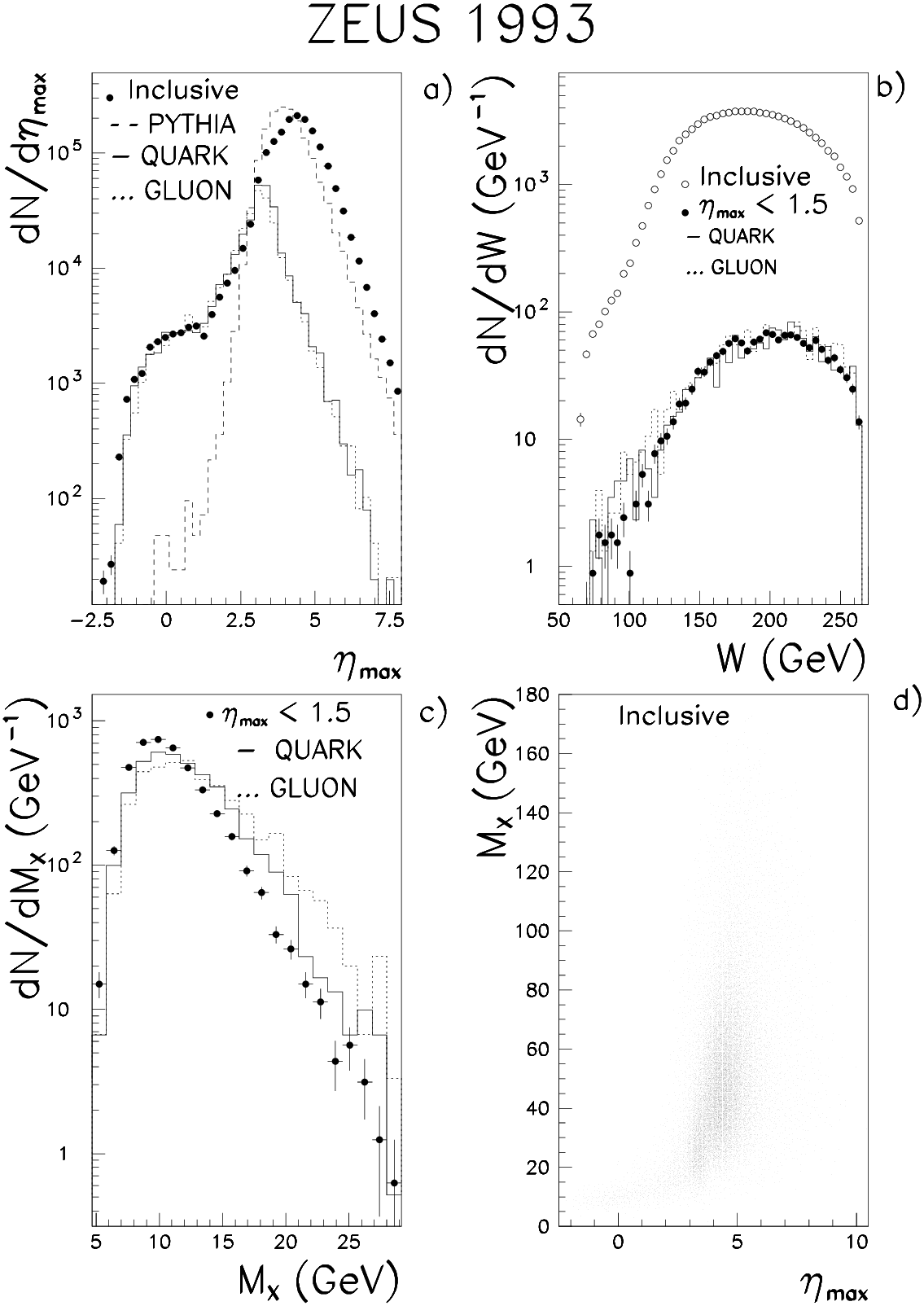}
\vspace{19cm}
\caption{
\newline
(a) The distribution of $\eta_{max}$, the pseudorapidity of the most
forward calorimeter condensate with energy above 400 MeV for
the photoproduction sample with $E_T > 5$ GeV
along with the predictions from PYTHIA (dashed line) and POMPYT with
a quarkonic (solid line) or gluonic (dotted line) pomeron.
\newline
(b)  The distribution in $W$ for all events and for those with
$\eta_{max}\leq 1.5$. The latter are well described by pomeron induced
reactions as implemented in POMPYT.
\newline
(c) The mass of the hadronic system $M_X$ for events with a large
rapidity gap as defined by $\eta_{max}\leq 1.5$ along with the
POMPYT predictions.
\newline
(d) A scatter plot of the
mass of the hadronic system, $M_X$, versus $\eta_{max}$.}
\end{figure}

% fig 2

\newpage
\clearpage
\begin{figure}[h]
\epsfysize=10cm
\includegraphics{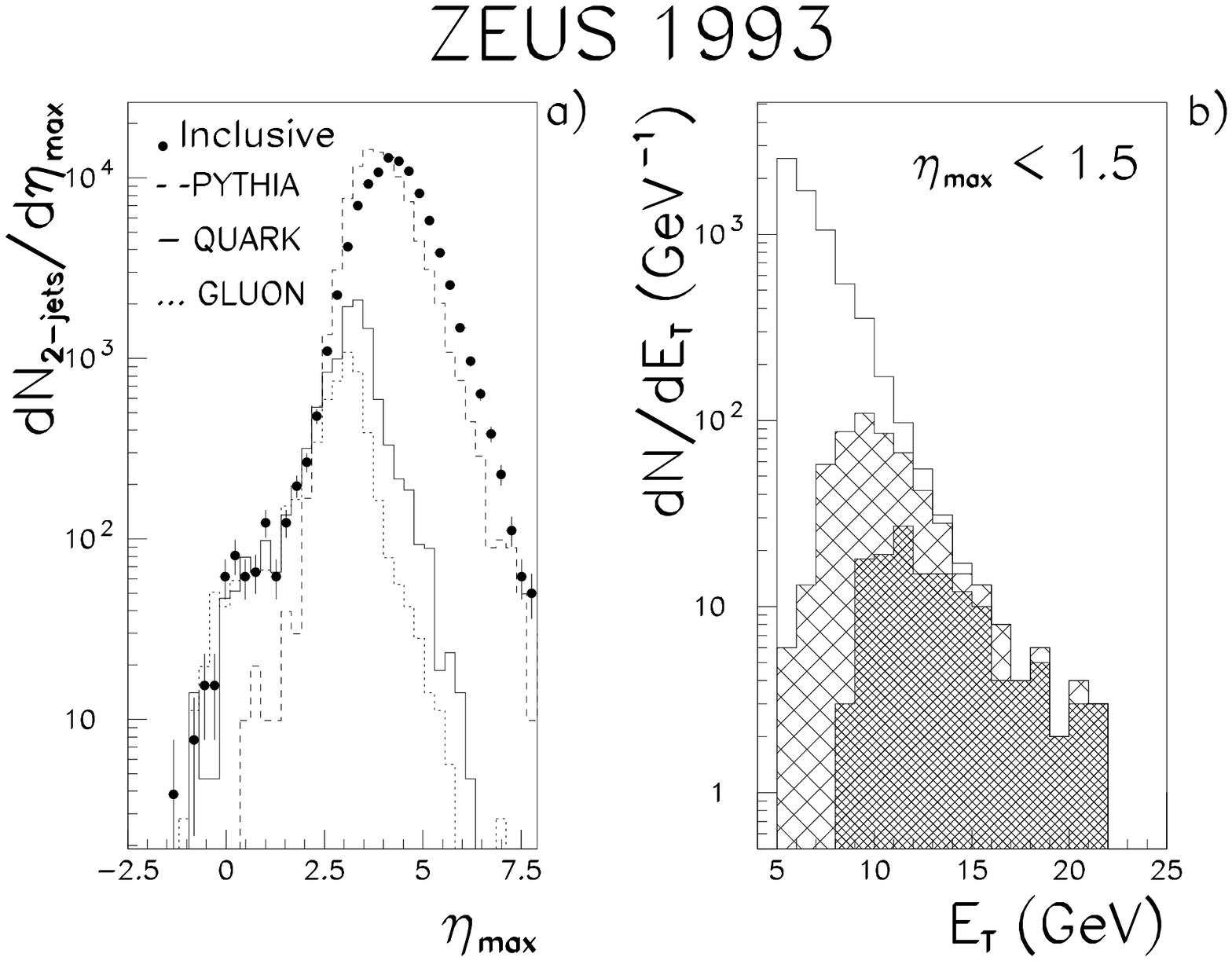}
\includegraphics{suceso.v1.ps}
\vspace{19cm}
\caption{
\newline
(a) The distribution of $\eta_{max}$ in the photoproduction sample
with $E_T > 5$ GeV and two or more jets
along with the predictions from PYTHIA (dashed line) and POMPYT with a
quarkonic (solid line) or gluonic (dotted line) pomeron.
\newline
(b) The distribution of the total transverse
energy $E_T$ for the photoproduction sample with a large rapidity gap
and, in addition, for the subsample of those
events with at least one (cross hatched area) and at
least two (shaded area) jets in the final state.
\newline
(c) A display of a
two-jet event in hard photoproduction with a large rapidity gap.}
\end{figure}

\newpage
\begin{figure}[h]
\includegraphics{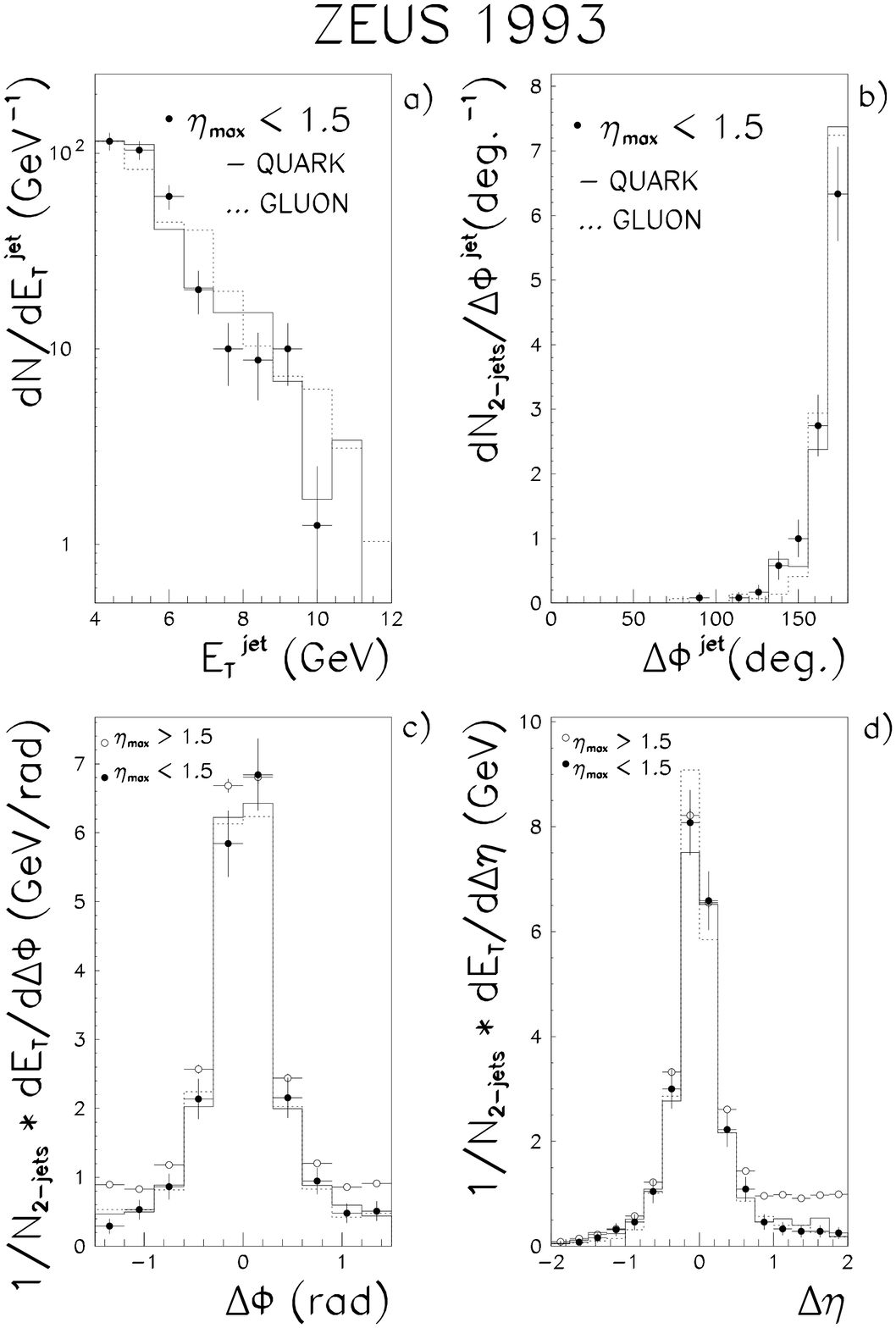}
\vspace{19cm}
\caption{
\newline
(a) The transverse
energy distribution of the jets in the two-jet sample in photoproduction
with $\eta_{max} \leq 1.5$ along with the POMPYT predictions for a
quarkonic (solid line) or gluonic (dotted line) pomeron.
\newline
(b) The $\Delta \phi^{jet}$ distribution for two-jet
events in photoproduction with $\eta_{max} \leq 1.5$ along with a
comparison with the POMPYT predictions as in (a).
\newline
(c, d) The transverse energy weighted profiles for jets with
 $\eta_{jet} < 0$  in photoproduction
events with $E_T > $ 5 GeV together with POMPYT predictions as in (a).
Two samples are presented: one with
$\eta_{max} \le $1.5 and the other with $\eta_{max} > 1.5$; (c) shows
the azimuthal profiles and (d) shows the pseudorapidity profiles.}
\end{figure}

\newpage
\begin{figure}[h]
\includegraphics{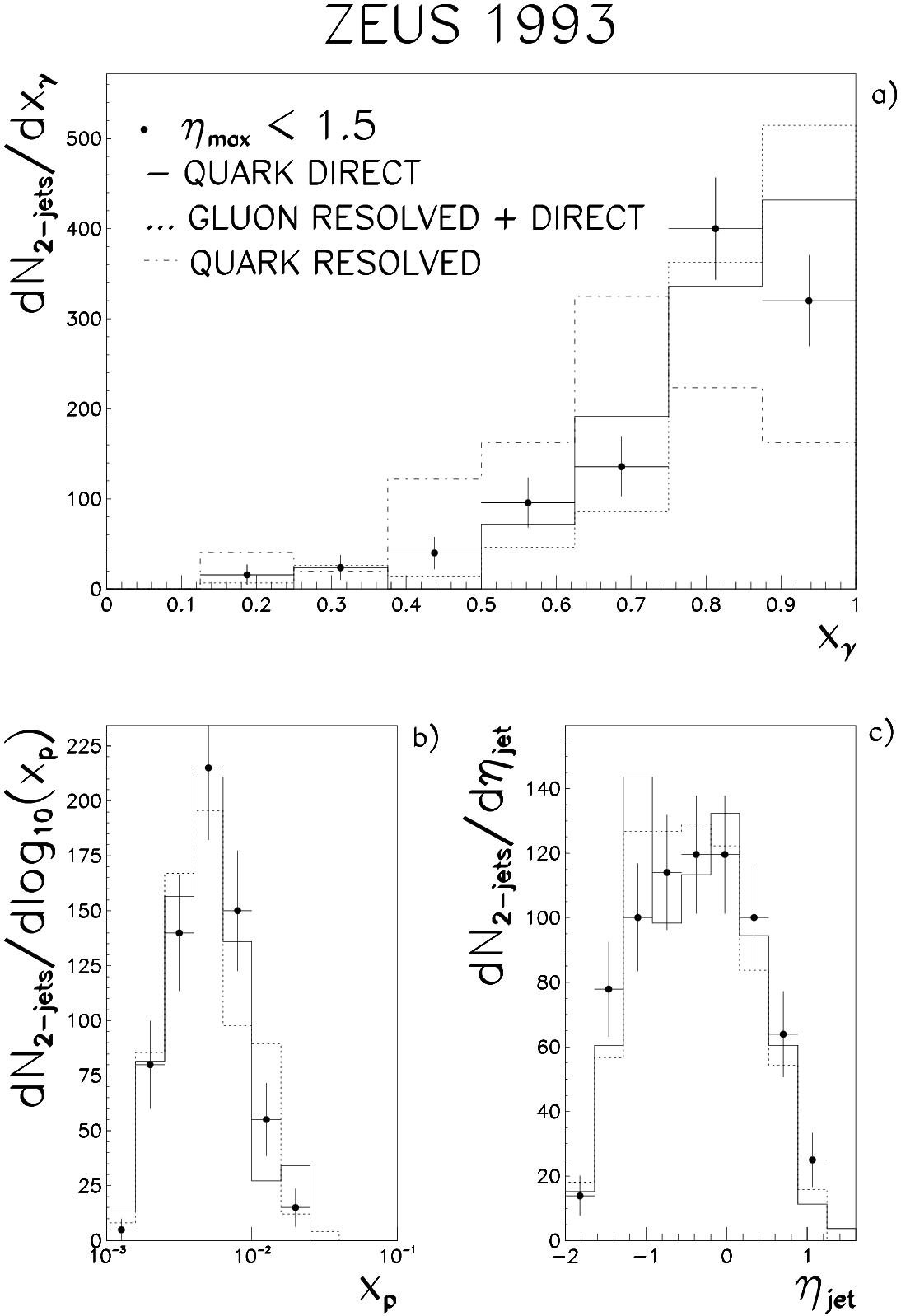}
\vspace{19cm}
\caption{
\newline
(a) The
$x_\gamma$ distribution for
photoproduction events with a large rapidity gap (dots) compared to
POMPYT predictions. The solid line represents the direct photon interaction
with a quarkonic  pomeron. The dotted line denotes the sum of the
direct and resolved interactions of the photon with a gluonic pomeron.
Also shown are the results
from POMPYT with a quarkonic pomeron using only the resolved photon processes
(dash-dotted  histogram). Each histogram has been normalised separately to the
number of data events.
\newline
(b) The measured fraction of the proton momentum, $x_p$,
entering the hard scattering for two-jet events, along with the POMPYT
predictions
for a quarkonic (solid line) or gluonic (dotted line) pomeron.
\newline
(c) The jet pseudorapidity distribution for the two-jet
sample and for the POMPYT predictions, as in (b).}
\end{figure}
\end{document}